*2024-2025 CRA Quadrennial Paper*

# Enabling the AI Revolution in Healthcare

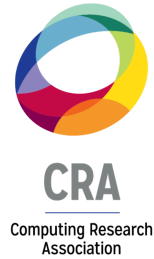
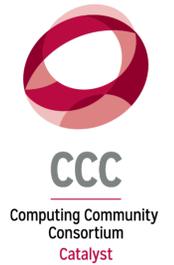

Mona Singh (Princeton University), Katie Siek (Indiana University Bloomington), David Danks (University of California San Diego), Rayid Ghani (Carnegie Mellon University), Haley Griffin (CRA), Brian LaMacchia (MPC Alliance), Daniel Lopresti (Lehigh University), Tammy Toscos (Parkview Health)

**The transformative potential of AI in healthcare — including better diagnostics, treatments, and expanded access — is currently limited by siloed patient data across multiple systems. Federal initiatives are necessary to provide critical infrastructure for health data repositories for data sharing, along with mechanisms to enable access to this data for appropriately trained computing researchers.**

Artificial Intelligence holds the promise of transforming medical care. By leveraging each individual's specific characteristics, environment, and medical history, AI is poised to guide clinicians in performing better risk assessments, making earlier diagnoses, and delivering precision treatments — thereby leading to improved health outcomes for all Americans. Automated AI tools will expand medical access in rural areas with a scarcity of physicians and contribute to lower total health care costs.

Federal initiatives to build health data repositories (e.g., by extending the Trusted Exchange Framework and Common Agreement™ (TEFCA™)) will fundamentally change how healthcare systems handle and share patient data. However, for AI to reach its full potential in healthcare, computing researchers must be able to access these data repositories after receiving appropriate training.

Since effective AI systems for healthcare will need clinical, environmental and lifestyle data across large cohorts of diverse individuals, we recommend:

1. **Investment in research infrastructure** to build secure systems that enable authorized users to access repositories of anonymized patient health data collected in clinics nationwide, including relevant contextual information, and prepared for computation.
2. **Regulatory changes** that support the democratization of anonymized and curated health data managed by electronic health record companies while providing incentives to health systems to partner with researchers to ensure health data are properly contextualized.

## Challenges

**AI is powered by data**. AI relies on machines learning from examples and subsequently making decisions or recommendations, sometimes as well as or even better than humans. AI methods process massive amounts of data — far more than any individual can — and uncover patterns within them. For example, given numerous samples of brain imaging data labeled as either exhibiting brain tumors or not, computers can use deep learning to identify the patterns that indicate cancer. Once "trained" in this manner, the computer can recognize these diagnostic patterns in new images much faster than a physician. In general, the larger and more diverse the datasets on which a given model is trained, the better it will perform on new data.

Although the United States has invested in accessible health-oriented data sets (e.g., *All of Us*, which currently includes genomic and clinical data for nearly one million participants), much more representative data is needed to create personalized medicine for all Americans. The current datasets are not large enough to uncover medically relevant patterns in individuals with infrequently observed combinations of symptoms or conditions. There is every reason to believe that the lesson of generative AI — from ChatGPT to DALL-E — where training on more data has led to dramatically better results, is equally applicable to medical applications of AI. As we innovate AI for healthcare data, we must ensure that quality data is ethically sourced from individuals by following established guidelines and standards (e.g., the Coalition for Health AI's *Assurance Standards Guide*).

**Digital medical data abounds, but has limited access**. In theory, AI can learn from clinical data collected nationwide. Each year in the U.S., there are more than one billion doctor's visits for over 100 million different patients. Health data from these interactions is stored in digital formats as electronic health records (EHRs). However, an individual's healthcare data is often fragmented across multiple institutions — from doctors' EHRs to radiology centers to groups that facilitate on-site vaccinations.

If an AI system were trained on EHR data across hospital systems — consisting of tens of billions (or more) of individual data points — it would have far more experience than any single doctor. Such an AI system could provide clinicians with recommendations for each patient by incorporating all known information about that patient and leveraging the characteristics, treatments, and health outcomes of patients across the country. Even an AI system that could retrieve diagnoses and outcomes for individuals with similar profiles to a given patient would be of enormous value to clinicians. As more heterogeneous patient data (e.g., genomic, environmental, or wearable device data) is collected and shared, AI systems trained on these data could enable highly individualized treatments, making precision medicine efficiently available to all.

Unfortunately, it is challenging for any machine learning model to access the full breadth of this data. Individual hospital systems are often hesitant to share health data with external hospitals due to business-related concerns. Some software companies that manage EHRs have created proprietary datasets from health systems using their software (e.g., Epic's Cosmos, which has data from nearly 1,600 hospitals and 36,000 clinics representing 277 million patients) and have built recommendation systems based on these data. However, these datasets cover only a subset of patients and are not available to researchers who are unaffiliated with these health systems. This approach can stifle innovation, keeps healthcare costs high, and prevents the full benefits of AI reaching the field of medicine.

The lack of a comprehensive, accessible, representative healthcare dataset is the greatest obstacle to creating revolutionary AI systems for medicine. Incentives are needed from the federal government (e.g., mandates from the Office of the National Coordinator and Health and Human Services) to ensure that EHR and other medically relevant data are made "AI-ready" and easily shared. Researchers and other stakeholders who can improve healthcare systems and delivery must be granted access to these data. Just as the availability of text and image data across the internet has led AI researchers to focus on these domains — yielding stunning advances — the availability of large medical datasets would drive increased focus by AI and health researchers on critical problems in healthcare.

## Recommendations

**Investing in secure infrastructure for making health data accessible while respecting privacy rights.** Unlike image or text data, personal healthcare data cannot reside on the internet in a fully accessible system. It is necessary to set up a system that is secure and preserves individual privacy. We acknowledge the work on the Trusted Exchange Framework and Common Agreement™ (TEFCA™) as a step toward interoperability; however, more clarity is needed in terms of access, privacy, and patient participation. Additionally, as currently implemented, TEFCA™ provides a means for sharing data about specific patients, but its goals would need to be extended to build the types of data repositories — whether centralized or distributed — necessary for training AI systems.

In addition to requiring these data storage systems to be HIPPA-compliant, it is essential that anyone using the data to train AI systems must not have access to information that could be used to identify the individuals whose data they are using. One approach to preserving patient privacy is to separate all identifying information from the health records added to research databases. This approach is used by the National Institutes of Health (NIH) in its *All of Us* Research Program, which shares genomic and clinical data in a de-identified format. Extending and scaling up such efforts to handle EHRs and other medical data for all patients in the United States should be a top priority, given the importance of healthcare to every American.

Alternate approaches, such as OpenSAFELY, keep data distributed across healthcare systems but accessible via specific, logged, auditable queries to a central system that grants access to highly controlled subsets of data and/or aggregated data. Another approach to preserving patient privacy may be through advanced encryption techniques such as fully homomorphic encryption (which allows computation on encrypted data) and secure multiparty computation (which enables multiple parties to jointly compute on data without revealing privately held values).

An additional challenge is that these solutions primarily focus on clinical data, while social determinants of health — which account for 80 percent of health outcomes — must also be considered. Privacy-preserving and cryptographic approaches, including secure and private machine learning and unlearning, would benefit from additional research funding to explore their application to public health data analysis.

**Invest in research on how people can effectively have agency over their own health data.** Not only will we need to develop technology to protect privacy when sharing data, but we must also create mechanisms that allow patients to control whether and under what circumstances their data may be shared. Patients may not want their provider or healthcare system to use, access, or share their health data for purposes beyond their own treatment. Conversely, some patients may prefer that their data is shared only for specific purposes (e.g., a clinical trial) or with designated groups (e.g., researchers).

The onus should be on care providers to ask permission to access a patient's data and be transparent about how they would use it — not the other way around. There is an urgent need for research into the secure and explainable handling and sharing of health data. This work could broadly benefit all Americans, as CRA's 2020 Quadrennial Paper, *Modernizing Data Control*, advocated for individuals to *"control the use of their data, extract value from their data, and understand their exposure to harms."*

**Incentivize EHR companies to share data**. Epic's Cosmos and TEFCA™ are promising mechanisms for computing researchers to access health data and develop novel AI models for cost-efficient personalized medicine. However, a major impediment is that very few computing researchers can have easy access to data from these systems. Mechanisms should be established to ensure that all researchers, regardless of their affiliations with specific healthcare facilities, can undergo appropriate training and gain access to these datasets.

For example, in its All of Us research program, the NIH has developed a model in which researchers can undergo training, register their research projects, and then access data through controlled channels. Additionally, financial incentives —similar to those used to encourage the adoption of EHRs — may be necessary to encourage healthcare providers to share data, ensuring a more comprehensive and holistic view of individuals' health journeys.

**Support research for necessary technical innovations**. To realize the data sharing necessary to build accurate AI systems for healthcare, numerous innovations are required in fields such as databases, security and privacy research, scalable computer systems, computational biology, artificial intelligence, fairness in machine learning, sociotechnical system design, and ethics.

- **Data standardization and translation schemas** of EHRs from different healthcare organizations are essential to ensure interoperability.
- **Medical AI systems must be designed and tested with fairness in mind** to identify and address existing disparities in the healthcare system. This includes modeling for large populations of Americans who currently lack access to healthcare.
- **Infrastructure must be developed to share health data broadly** in a privacy-preserving way while enabling the development of cutting-edge AI models.

The realization of AI's full potential in medicine depends on ensuring that researchers have the necessary tools, data, and frameworks to innovate while safeguarding patient privacy and maintaining public trust.

---

*This quadrennial paper is part of a series compiled every four years by the **Computing Research Association (CRA)** and members of the computing research community to inform policymakers, community members, and the public about key research opportunities in areas of national priority. The selected topics reflect mutual interests across various subdisciplines within the computing research field. These papers explore potential research directions, challenges, and recommendations. The opinions expressed are those of the authors and CRA and do not represent the views of the organizations with which they are affiliated.*

*This material is based upon work supported by the U.S. National Science Foundation (NSF) under Grant No. 2300842. Any opinions, findings, and conclusions or recommendations expressed in this material are those of the authors and do not necessarily reflect the views of NSF.*